\documentclass[10pt,letterpaper]{article}
\usepackage[top=0.85in,left=2.75in,footskip=0.75in]{geometry}

\usepackage{amsmath}  
\usepackage{amssymb}  
\usepackage{pifont}

\usepackage{changepage}

\usepackage{textcomp,marvosym}

\usepackage{cite}

\usepackage{nameref,hyperref}

\usepackage[right]{lineno}

\usepackage{pdfpages}

\usepackage[nopatch=eqnum]{microtype}
\DisableLigatures[f]{encoding = *, family = * }

\usepackage[table]{xcolor}

\usepackage{array}
\usepackage{pdfpages}
\usepackage{caption}   
\usepackage{hypcap}    

\newcolumntype{+}{!{\vrule width 2pt}}

\newlength\savedwidth

\usepackage{bm}
\usepackage{comment}

\usepackage{float}

\raggedright
\usepackage[aboveskip=1pt,labelfont=bf,labelsep=period,justification=raggedright,singlelinecheck=off]{caption}

\makeatletter
\renewcommand{\@biblabel}[1]{\quad#1.}
\makeatother

\usepackage{lastpage,fancyhdr,graphicx}
\usepackage{epstopdf}
\fancyheadoffset[L]{2.25in}
\fancyfootoffset[L]{2.25in}
\begin{document}
\vspace*{0.2in}

\begin{flushleft}
{\Large
\textbf\newline{Circadian output network can buffer period variability} 
}
\newline
\\
Ismail M Nur\textsuperscript{1},
Hotaka Kaji\textsuperscript{1},
Yuzuru Mitsui\textsuperscript{1},
Hiroshi Ito\textsuperscript{1,*}
\\
\bigskip
\textbf{1} Faculty of Design, Kyushu University, Fukuoka, Japan
\\
\bigskip

%
%





* hito@design.kyushu-u.ac.jp

\end{flushleft}
\section*{Abstract}
Circadian rhythms are biological oscillations that govern 24-hour physiological and behavioral processes across most organisms. Recent bioimaging studies have revealed that even individual cells can exhibit circadian rhythms. The period of cellular oscillations can fluctuate due to molecular noise in the circadian clock machinery. Whether regulatory networks downstream of the clock amplify or attenuate clock-derived period fluctuations remains poorly understood. In this study, we numerically observed period variability in a self-sustained oscillator coupled to an output network. Our numerical calculations demonstrated that a serial pathway does not merely relay timing signals but actively shapes rhythmic reliability. The extent of this reduction depended on parameters of both the clock and output systems. For more complex output networks, the shortest-path length from the core oscillator was a major determinant of increased oscillation precision. This noise-buffering effect saturated in long cascades. These results suggest the existence of an intrinsic precision-enhancing mechanism embedded within circadian output networks.

\section*{Author summary}
Many organisms, from bacteria to humans, rely on circadian clocks to coordinate daily activities such as sleep, metabolism, and hormone production. Although these clocks maintain remarkably regular timing, they are composed of molecular components that operate in noisy cellular environments. As a result, the timing of individual cellular clocks can fluctuate. How biological systems maintain reliable daily rhythms despite this noise remains an important open question.

In this study, we used mathematical models to investigate how circadian timing information is transmitted from a circadian clock through downstream output pathways. We found that these pathways do more than simply pass timing signals from one component to another. Instead, they can reduce fluctuations in transmitted timing signals and improve the reliability of rhythmic outputs. In particular, timing signals became progressively more precise as they traveled through multi-step pathways. This effect was observed across different network architectures and was closely associated with the network distance between a downstream component and the core clock. However, the improvement in precision eventually saturated as pathway length increased. Our findings suggest that circadian output pathways may contain a built-in mechanism for enhancing timing precision to help organisms maintain robust daily rhythms in the presence of molecular noise.


\section*{Introduction}
Circadian rhythms are biological oscillations that repeat approximately every 24 hours. They govern a wide range of physiological and behavioral processes in most organisms, including mammals, plants, fungi, and cyanobacteria. The circadian machinery comprises three parts: the central clock that generates self-sustained oscillations, the input system that links environmental day-night cycles to the clock, and the output system that transmits circadian timing information to physiological processes. A defining feature of the central circadian clock is self-sustained rhythmicity: it persists even under constant environmental conditions without external cues \cite{Johnson:2004book}. Single-cell studies have shown that individual, uncoupled cells, such as those in dissociated cultures, retain robust circadian oscillations \cite{nagoshi2004circadian, mihalcescu2004resilient, nakamura2022adaptive}. These observations establish that circadian timekeeping is an intrinsic cellular property and does not require intercellular coupling to sustain rhythmicity.

Despite their regularity, circadian rhythms show period fluctuations due to transcriptional noise \cite{gonze2002robustness, nishino2013transcription} and environmental fluctuations such as irregular light, temperature, and feeding patterns \cite{micklem2021cut, chabot2007stochastic}. This variability is commonly measured by the coefficient of variation (CV), the ratio of the standard deviation to the mean period. For example, mammalian cells and \textit{Arabidopsis} seedlings show similar period variability $ \text{CV} \approx 0.046 $, while cyanobacteria exhibit much higher precision $ \text{CV} \approx 0.005 $ \cite{mihalcescu2004resilient, gould2018coordination, li2020noise}. While the molecular origins of such differences in precision have been extensively studied at the level of the core circadian oscillator, how this variability is reshaped downstream of the clock remains less well understood.

Noisy oscillatory signals generated by the central clock propagate into the output system and can thereby shape period variability in circadian outputs. The output system typically consists of multilayered molecular networks that regulate diverse physiological processes, including gene expression, protein modifications, and metabolism. 

Importantly, different circadian outputs often exhibit distinct waveform shapes and levels of period variability, even within the same cell type. However, it has not been experimentally established whether output networks generally amplify or attenuate period fluctuations. Moreover, it remains unclear whether specific molecular circuit architectures systematically control the variability of circadian periods, or whether output precision simply reflects that of the core oscillator.

These unresolved questions highlight the importance of regulatory network structure in shaping stochastic signal transmission. Previous studies have shown that gene regulatory architecture can strongly influence noise dynamics and output stability. For example, negative autoregulation reduces fluctuations and accelerates response times \cite{becskei2000engineering,dublanche2006noise}, while broader analyses demonstrate that circuit topology can determine whether noise is buffered, amplified, or propagated through regulatory layers \cite{chalancon2012interplay}. Beyond simple feedback, feed-forward loops have been shown to implement temporal signal-processing operations such as delay, pulse generation, and noise filtering \cite{mangan2003structure}. Together, these findings suggest that variability is not solely a property of individual molecular components but is fundamentally shaped by network topology, motivating theoretical approaches that explicitly examine how noise propagates in a structured circadian output network.

Theoretical studies have explored circadian variability using both dynamical and statistical approaches. Period variability has been analyzed in coupled oscillator systems \cite{kori2012structure, mori2013period}, and deviations in period have been exploited to infer underlying network properties \cite{mori2022noninvasive}. Mori and Mikhailov (2016) demonstrated that rhythm accuracy depends on the variable being observed, implying that different molecular components within the same oscillator can display distinct levels of precision \cite{mori2016precision}. Extending this idea, Kaji et al. (2023) analyzed a minimal two-node model consisting of a clock and a downstream output, showing that period fluctuations at the output can be smaller than those of the clock itself \cite{kaji2023enhanced}. This result indicates that downstream systems can enhance rhythmic precision without requiring explicit noise-filtering mechanisms. Nevertheless, whether such precision enhancement persists in networks with multiple outputs and structured coupling architectures remains an open question.

In this study, we extended the theoretical framework of Kaji et al. (2023) \cite{kaji2023enhanced} to circadian systems with multiple outputs. Using a mathematical model of a noisy circadian oscillator coupled to three output nodes, we systematically investigated how waveform shape, coupling structure, and the location of noise sources influenced period variability across different outputs. By focusing on a minimal yet generalizable model, we aimed to identify fundamental principles that govern how downstream networks attenuate or amplify circadian period variability. Our results provide testable predictions for single-cell experiments and suggest that the precision of circadian rhythms is not solely a property of the core clock but can emerge from the architecture and dynamics of the output system.

\section*{Model}

We considered a fluctuating circadian oscillator coupled to an output system to investigate how the architecture of the output network affects rhythmic precision. We adopted a self-sustained oscillator in the presence of noise as a circadian clock:

\begin{equation}
\frac{1}{T}\dot{\bm{x}} = \bm{f}(\bm{x})
 + \epsilon\sqrt{D}\boldsymbol{\xi}(t),
\label{eq:unified_general}
\end{equation}
where $\bm{x}$ represents the state vector of the clock. Eq. \eqref{eq:unified_general} admits a limit-cycle solution when $\epsilon=0$, i.e., in the absence of noise. $T$ is the oscillation period of $\dot{\bm{x}}=\bm{f}(\bm{x})$ and acts as a scaling parameter to normalize the oscillation period of Eq. \eqref{eq:unified_general} to unity. $\bm{\xi}(t)$ represents noise applied to the clock system. $D$ denotes the noise intensity.  

The oscillatory signal generated by the circadian clock is transmitted to the output system, whose state vector is denoted by $\bm{y}$. Its dynamics is described by the linear differential equation:
\begin{equation}
\dot{\bm{y}} = (\beta A-\gamma I)\bm{y} + \beta B\bm{x}+\bm{\alpha},
\label{eq:output}
\end{equation}
where $A$ and $B$ are the adjacency matrices representing connectivity within the output system and from the clock to the output system, respectively, i.e., the $(i,j)$ entry is 1 if there is a connection from $j$ to $i$, and 0 otherwise. $I$ denotes the identity matrix. We assumed, for simplicity, that the coupling and degradation rates were identical for all output variables, with values $\beta$ and $\gamma$, respectively. $\bm{\alpha}$ represents the vector of basal synthesis rates.

\section*{Results}

\subsection*{Output pathway enhances circadian precision}

We first considered the stochastic Goodwin model \cite{gonze2002robustness, goodwin1965oscillatory} coupled to a downstream output system (Fig. 1A). The Goodwin model represents a negative feedback loop, which can be described as 
\begin{equation}
\bm{f}(\bm{x}) =
\begin{pmatrix}
k_u (K+w^h)^{-1} - d_u u \\
k_v u - d_v v\\
k_w v - d_w w
\end{pmatrix},\;\bm{\xi}(t)=\begin{pmatrix}\xi(t)\\0\\0
\end{pmatrix}
\end{equation}
where ${}^t\bm{x} =(u,v,w)$ and the superscript $t$ denotes transposition. In the context of circadian clocks, $u$, $v$, and $w$ are interpreted as the concentrations of mRNA of a clock gene, a clock protein, and the modified form of the clock protein, respectively \cite{kurosawa2002comparative}. $\xi(t)$ is independent Gaussian noise satisfying 
$E[\xi(t)]=0$, $E\left[\xi(t)\xi(t')\right]=\delta(t-t')$, where $E[\cdot]$ represents the expectation and $\delta(t)$ is the Dirac delta function. The parameters $k_*$, $K$, $d_*$, and $h$ represent synthesis rates, the repression threshold, degradation rates, and the Hill coefficient, respectively.

As for the output system, we assumed the simplest cascade consisting of three components, ${}^t\bm{y} = (y_1, y_2, y_3)$, in which the clock sequentially regulates three downstream variables along the pathway. In addition, we assumed that $w$ transmits the oscillatory signal to the output system. Thus, we analyzed the output system represented by the diagram, $w \rightarrow y_1 \rightarrow y_2 \rightarrow y_3$, i.e.,  Eq. \eqref{eq:output} can be written as
\begin{equation}
\begin{pmatrix}
\dot{{y}_1} \\
\dot{{y}_2} \\
\dot{{y}_3}
\end{pmatrix}
= \beta 
\begin{pmatrix}
0 & 0 & 0 \\
1 & 0 & 0 \\
0 & 1 & 0
\end{pmatrix}
\begin{pmatrix}
y_1 \\
y_2 \\
y_3
\end{pmatrix} + 
\beta 
\begin{pmatrix}
0 & 0 & 1 \\
0 & 0 & 0 \\
0 & 0 & 0
\end{pmatrix} 
\begin{pmatrix}
u \\
v \\
w
\end{pmatrix} - \gamma
\begin{pmatrix}
y_1 \\
y_2 \\
y_3
\end{pmatrix} + \begin{pmatrix}
\alpha_1 \\
\alpha_2 \\
\alpha_3
\end{pmatrix}.
\label{eq:output_detail}
\end{equation}
 This structure represents a minimal hierarchical transmission pathway in which temporal information propagates stepwise from upstream to downstream components. 

Eq. \eqref{eq:output_detail} is an extension of the previous study \cite{kaji2023enhanced}, which examined a minimal clock--output network and revealed that period variability can be buffered in the downstream layer within an appropriate range of the degradation rate of the output component, $\gamma$. This result suggests that oscillation precision may change across circadian output networks. However, it remains unknown whether, as the network depth increases, precision continues to improve monotonically downstream, or variability re-emerges at some layers. 

We numerically solved the Goodwin oscillator coupled to an output system and quantified the CV of the periods of both the circadian clock $\bm{x}$ and output system $\bm{y}$. The numerical simulation confirmed that the CV of the oscillation periods could decrease progressively from the clock signal $w(t)$ to $y_1(t)$, $y_2(t)$, and $y_3(t)$, indicating that deeper downstream cascades could further buffer timing fluctuations within the parameter ranges examined (Fig. 1B). However, this decrease of CV was not universal and depended on the degradation rate $\gamma$; the ordering of CV could reverse for other values of $\gamma$, i.e., a more distal node could exhibit a larger CV (Fig. S1). Notably, the amplitude was also variable with the positions of the nodes. These variations in amplitude may contribute to differences in period fluctuations.

\subsection*{Parameter-dependent propagation of period fluctuations}

To further examine the generality of the fluctuation-suppression effect by the output system, we focused on a coupled system consisting of a phase model, representing a general reduced form of a self-sustained oscillator, and its output network:
\begin{equation}
\begin{aligned}
 \dot{\theta}&=\omega +\epsilon \sqrt{D}\xi(t)\\
\dot{y_1} &= \alpha_1 + \beta \sin \theta -\gamma_1 y_1\\
\dot{y_2} &= \alpha_2 + \beta y_1 -\gamma_2 y_2\\
\dot{y_3} &= \alpha_3 + \beta y_2 -\gamma_3 y_3,
\end{aligned}
\label{eq:phase}   
\end{equation}
where $\theta(t) \in [0, 2\pi)$ denotes the phase variable of the clock system, $y_1, y_2$ and $y_3$ represent the components in the output system. A previous study reported that the degradation rate of downstream output molecules crucially determines fluctuation dynamics \cite{kaji2023enhanced}. Motivated by this fact, we analyzed the propagation of period fluctuations across three downstream layers in the output system for different values of the degradation-rate $\gamma$.

Numerical simulations of Eq. \eqref{eq:phase} showed that period fluctuations in all output variables exhibited a unimodal dependence on $\gamma$, characterized by a single local minimum (Fig. 2A). Both the minimum CV value and the corresponding $\gamma$ at which the minimum occurred were dependent on the depth of the variable in the output pathway. Thus, different CV orderings were observed, such as
$\mathrm{CV}[y_1]>\mathrm{CV}[y_2]> \mathrm{CV}[y_3]$ 
or
$\mathrm{CV}[y_3]>\mathrm{CV}[y_2]> \mathrm{CV}[y_1]$. In other words, fluctuations in the oscillatory signal can either be amplified or attenuated depending on the parameter value.

Interestingly, under the conditions in which the fluctuations of the component located at the lowest layer of the network were most strongly suppressed, signal propagation consistently exhibited a denoising effect. Specifically, at the value of $\gamma$ that minimized the fluctuations of $y_2$, we observed that $\mathrm{CV}[y_1] > \mathrm{CV}[y_2]$. Similarly, at the value of $\gamma$ that minimized the fluctuations of $y_3$, the ordering was $\mathrm{CV}[y_1] > \mathrm{CV}[y_2] > \mathrm{CV}[y_3]$.

These properties observed in the phase model were reproduced in the Goodwin model coupled to the output system. Furthermore, the same behavior was observed in the FitzHugh–Nagumo model (Fig. 2A), which captures activation and recovery in neural dynamics using a fast activator $v(t)$ and a slow inhibitor $w(t)$:
\begin{equation}
 \bm{f}(\bm{x}) =
 \begin{pmatrix}
v(a-v)(v-1) - w\\
b v-c w
\end{pmatrix},\;\bm{\xi}(t)=\begin{pmatrix}0\\ \xi(t)
\end{pmatrix}
\end{equation}
where ${}^t\bm{x} =(v,w)$; $a$, $b$, and $c$ are parameters, and $\xi(t)$ represents noise with the same stochastic properties as described above. These results suggest a general principle of the suppression of period variability in output-system networks that is independent of the oscillator's detailed structure.

In addition, deeper outputs tended to exhibit smaller minimum CV values, i.e., $\min_{\gamma}\mathrm{CV}[y_1] > \min_{\gamma}\mathrm{CV}[y_2] > \min_{\gamma}\mathrm{CV}[y_3]$. This ordering was also observed in the Goodwin and FitzHugh–Nagumo models when the system was close to a Hopf bifurcation, where self-sustained oscillations vanish (Fig. 2B). By contrast, the opposite ordering, $\min_{\gamma}\mathrm{CV}[y_3]
> \min_{\gamma}\mathrm{CV}[y_2]
> \min_{\gamma}\mathrm{CV}[y_1]$,
occurred over a larger region of parameter space (Fig. S3A). However, the ordering of CV values was not always stable across the entire degradation-rate range. While the ranking could vary for \(1 \leq \gamma \leq 10\), the relation $\mathrm{CV}[y_1]
> \mathrm{CV}[y_2]
> \mathrm{CV}[y_3]$ was consistently maintained for larger degradation rates (\(\gamma \geq 10\)). When a system is close to a Hopf bifurcation point, the system more closely resembles a harmonic oscillator. Thus, the waveforms generally approach a sinusoidal shape near the Hopf bifurcation (Fig. S3B, \cite{guckenheimer2013nonlinear}). In addition, waveform properties can modulate period variability \cite{kaji2025sinusoidal}. Taken together, these results suggest that waveform shape influences the ordering of minimum CV values across different degradation rates $\gamma$. These findings further indicate that clock-output coupled systems may share a general principle governing the parameter dependence of period variability.

\subsection*{Distance from the clock determines precision in the output system}

So far, we have examined the transmission of fluctuations in the circadian period through a downstream network composed of three serially connected components. Because circadian systems typically possess more complex output architectures, we next considered some extended networks in which additional lateral connections were introduced into the serial topology. Exclusion of wiring patterns with closed feedback loops or self-feedback yields eight admissible network topologies (Fig. 3A, Methods). We chose the system's parameter values that yielded a progressive reduction in fluctuations along the serial network, such that $\mathrm{CV}[y_1]>\mathrm{CV}[y_2]>\mathrm{CV}[y_3]$. Among the network architectures examined, the simple cascade structure most consistently produced the lowest period variability (CV) across the models (Goodwin, FitzHugh–Nagumo, and phase models). To further quantify this trend, we replotted the output CV as a function of the shortest-path distance from the clock, defined as the number of edges along the shortest path from the clock to each output (hereafter, the distance from the clock). This analysis revealed a clear monotonic relationship: outputs farther from the clock exhibited lower CV, whereas additional lateral connections did not enhance oscillation-period precision.

We further investigated a more complex output network: the Goodwin oscillator connected to all four possible four-component output networks with distinct topologies, rather than analyzing each output topology separately (Fig. 3B, Methods). Despite the increased complexity and highly branched topology, the same overall trend persisted: outputs farther from the clock generally exhibited lower period CV, with the most distal node (i.e., 4 edges) displaying the smallest variability. These findings further support the idea that distance from the clock is a robust predictor of noise-buffering efficiency.

In addition, we investigated whether the reduction in fluctuation of the oscillation period continues indefinitely as distance increases or eventually saturates. We therefore generated a longer sequential cascade network that connects to the Goodwin model and then evaluated period CV across the chain (Fig. 3C, Methods). We found that period CV decreased with distance initially but reached a minimum around the 10th output and then started to increase for further downstream output components, indicating that noise buffering has a finite effective range (Fig. 3C). We observed qualitatively similar trends in the long cascade connecting to FitzHugh–Nagumo and phase models, although the minimum CV was observed at different distances from the clock (Fig. S4).

\section*{Discussion}

This study demonstrated that circadian output pathways can buffer period variability, although the extent of buffering depends on both intrinsic oscillator parameters and network architecture. In addition, oscillation precision generally improved with increasing distance from the clock source. However, this benefit eventually saturated in sufficiently deep cascades. These findings suggest that the downstream circadian output network actively shapes rhythmic reliability. 

Physiological outputs from circadian clocks are typically regulated not by direct clock–output connections but by complex intermediate pathways \cite{dibner2010mammalian, mohawk2012central, takahashi2017transcriptional}. For example, in the mammalian adrenal steroidogenic pathway, the circadian clock controls the expression and activity of the rate-limiting steroidogenic acute regulatory protein (StAR) through multiple regulatory layers \cite{Son2008}. Likewise, in lipid metabolism, genome-wide transcriptional regulation mediated by REV-ERB$\alpha$ has been extensively characterized \cite{Cho2012}. In cyanobacteria, rhythmic phosphorylation of KaiC, the core component of the cyanobacterial clock, is transmitted through the SasA–RpaA two-component signaling system and regulates global transcriptional activity via RpaA \cite{Markson2013}. Our findings suggest that such indirect architectures may have evolved not only to diversify physiological outputs but also to improve temporal precision by increasing the effective distance from the core oscillator. Because different physiological functions may require different levels of temporal precision, the size and architecture of circadian output networks may have diversified to meet these distinct functional demands.

We found that rhythmic signals from phase oscillators and oscillators operating near a Hopf bifurcation, whose waveforms are approximately sinusoidal, tend to exhibit substantial attenuation of period fluctuations in downstream output networks. This finding is consistent with a previous study reporting that rhythms with more sinusoidal waveforms are less susceptible to fluctuation propagation \cite{kaji2025sinusoidal}. Notably, circadian clocks often approach a Hopf bifurcation under low-temperature conditions and consequently display nearly sinusoidal oscillations \cite{murayama2017low, xiao2025low}. Even at physiological temperatures, circadian rhythms generally exhibit smoother and more sinusoidal waveforms than the pulsatile activity patterns commonly observed in neuronal systems. From the perspective of fluctuation attenuation, these observations raise the possibility that sinusoidal oscillations may represent an adaptive feature of circadian systems. Such waveforms may facilitate the transmission of temporally precise rhythmic information through downstream regulatory networks while minimizing the amplification of stochastic fluctuations.

According to control theory \cite{delvecchio2014biomolecular}, the dynamical system $\dot{x}=u(t)-\gamma x$ acts as a first-order low-pass filter. Thus, the linear output system described in Eq. \eqref{eq:output} is expected to exhibit a similar low-pass property, which may help attenuate period fluctuations. However, period fluctuations are not always reduced and can, under certain conditions, even be amplified. This observation suggests that mechanisms beyond simple linear filtering may be involved in determining the precision of rhythmic outputs. Future work applying the analytical framework developed by Mori and Mikhailov \cite{mori2016precision} to Eqs.~\eqref{eq:unified_general}--\eqref{eq:output}, or to more general output networks with nonlinear regulatory interactions, may clarify the mechanisms underlying the attenuation of period fluctuations in circadian output networks.

\section*{Methods}

\subsection*{Measurement of CV in oscillation periods and peak detection}

We numerically solved Eqs.~\eqref{eq:unified_general}--\eqref{eq:phase} using the Euler--Maruyama method with a time step $\Delta t = 1.0 \times 10^{-4}$. To eliminate the influence of initial conditions, we discarded transient dynamics before analysis. Oscillation periods for both the core oscillator and downstream outputs were quantified using peak-to-peak intervals. Peaks in the time series were identified as local maxima using the \texttt{argrelmax} function from SciPy with \texttt{order = 100}. To exclude exceptional CV values caused by failures in peak detection, we calculated the mean CV and its standard error using only runs whose CV values fell within $1.5\sigma$ of the CV distribution across independent runs.

\subsection*{Expression of wiring patterns}
The connectivity matrices \(A\) and \(B\) in Eq.~\eqref{eq:output} specify the wiring pattern in the downstream network. For Fig. 3A, matrix \(A\) was restricted to a lower-triangular form, 
\[
A =
\begin{pmatrix}
0 & 0 & 0 \\
1 & 0 & 0 \\
a_{31} & 1 & 0
\end{pmatrix},
\]
where \(a_{31} \in \{0,1\}\), which constrains a feedforward organization without self-activation. The matrix \(B\) was restricted to
\[
B =
\begin{pmatrix}
0 & 0 & 1 \\
0 & 0 & b_{23} \\
0 & 0 & b_{33}
\end{pmatrix} (\mathrm{Goodwin}),
\quad
B =
\begin{pmatrix}
0 & 1 \\
0 & b_{22} \\
0 & b_{32}
\end{pmatrix}~\text{(FitzHugh--Nagumo}),
\]
\[
B =
\begin{pmatrix}
1 \\
b_{21} \\
b_{31}
\end{pmatrix} (\mathrm{phase\;model}),
\]
where all variable entries take binary values in \(\{0,1\}\). The three possible connections in each model give $2^3 = 8$ distinct network topologies displayed in Fig. 3A. For all three oscillator models, we used parameter sets near a Hopf bifurcation for which the ordering $\mathrm{CV}[y_1] > \mathrm{CV}[y_2] > \mathrm{CV}[y_3]$ was stable. 

For Fig. 3B, $A$ and $B$ are the union of all admissible four-node feedforward connectivity matrices. When merging two distinct networks, identical upstream subnetworks were unified into a single shared structure, whereas their downstream portions were retained as separate branches. By iteratively applying this procedure to all 64 networks, we constructed a single large network containing 64 downstream branches. The resulting network contained 75 output nodes distributed across discrete topological distances from the clock: 38 nodes at distance 1, 30 at distance 2, 6 at distance 3, and 1 at distance 4.

\section*{Supporting information}
\paragraph*{Supplemental figure 1}
Fig. S1. Different CV orderings across layers under different degradation rates. (PDF)
\paragraph*{Supplemental figure 2}
Fig. S2. Different orderings of minimum CV values across layers at degradation rates that minimize CV
($\min_{\gamma}\mathrm{CV}$). (PDF)
\paragraph*{Supplemental figure 3}
Fig. S3. Ordering of minimum CV values from proximal to distal outputs controlled by the repression threshold $K$. (PDF)
\paragraph*{Supplemental figure 4}
Fig. S4. Minimum CV values across distances from the clock. (PDF)

\section*{Author contributions}
\paragraph*{Conceptualization:} Ismail M Nur, Hiroshi Ito. 
\paragraph*{Investigation:} Ismail M Nur, Hotaka Kaji, Yuzuru Mitsui, Hiroshi Ito. 
\paragraph*{Methodology:} Ismail M Nur, Hotaka Kaji, Hiroshi Ito. 
\paragraph*{Supervision:} Hiroshi Ito.
\paragraph*{Visualization:} Ismail M Nur, Hiroshi Ito.
\paragraph*{Writing -- original draft:} Ismail M Nur, Hiroshi Ito.
\paragraph*{Writing -- review and editing:} Ismail M Nur, Hiroshi Ito.

\section*{Acknowledgments} 
We thank Professors F. Mori and M. Seki (Kyushu University) for fruitful discussions. 

\section*{Data availability statement}
All code used in this study is freely available on GitHub at https://github.com/hito1979/outputnoise/tree/main/arXiv2026
\section*{Funding}
This work was supported by the Japan Society for the Promotion of Science (JSPS) KAKENHI to H.I. (JP23H04475, JP25H02463), Y.M. (26K21337), AMED CREST to H.I. (JP24gm2010005), a Grant-in-Aid for JSPS Fellows to H.K. (JP24KJ1795), a JSPS Fellowship to Y.M. (26KJ0246), and a scholarship from the Japanese Government (MEXT) to I.M.N. (230526). The funders had no role in the study design, data collection and analysis, decision to publish, or manuscript preparation.
\section*{Competing interests}
The authors have declared that no competing interests exist.
\clearpage

\paragraph*{Fig. 1.}
\label{Fig. 1}
{\bf Reduced period fluctuations in downstream outputs from a clock.}
\textbf{A:} Schematic representation of a circadian clock system with a serially connected output network. The clock system generates oscillatory signals perturbed by noise. The output system processes these signals through multiple layers, with each layer subject to degradation and transcriptional regulation. Thus, timing signals propagate through the cascade. 
\textbf{B:} Time course of each variable of Goodwin model clock oscillator and output system. The parameter values were set to $D = 1.0 \times 10^{-3}$, $h = 10$, $k = 1$, $K = 1$, $T = 39.7$, $\epsilon = 0.01/T$, $d_u = d_v = d_w = 0.1$, $\alpha = 1$, $\beta = 11.5$, and $\gamma = 10.00$. Displayed CV values represent the mean across 100 repetitions (see Methods).

\paragraph*{Fig. 2.}
\label{Fig. 2}
{\bf Parameter dependence of period variability.}
\textbf{A:} Fluctuations in the oscillation period of the clock and the output variables in the Goodwin model, phase model, and FitzHugh–Nagumo model under different degradation rates. For the Goodwin model, the parameters were set to $k = 1$, $K = 100$, $\alpha_1 = \alpha_2 = \alpha_3 = 1$, $\beta = 1$, $h = 10$, $T = 39.7$, $\epsilon = 0.01/T$, $d_u = d_v = d_w = 0.1$, and $D = 1.0 \times 10^{-5}$. For the phase model, $\omega = 2\pi$, $D = 1.0 \times 10^{-2}$, and $\epsilon = 0.01$, with fixed $\alpha = 1$ and $\beta = 1$. For the FitzHugh–Nagumo model, $a = 0.5$, $b = 1.0$, $c = -0.5$, $\alpha = 1$, $\beta = 1$, $T = 7$, $\epsilon = 0.01/T$, and $D = 1.0 \times 10^{-5}$. Displayed CV values represent the mean of 100 repetitions (see Methods). 
\textbf{B:} Distribution of the regime
(stars) near a Hopf bifurcation in the parameter space of the Goodwin and FitzHugh–Nagumo oscillators. The background color represents the oscillation amplitude of the clock signal \(w\). For the amplitude maps (color scale), parameters were fixed at \(D = 0\), \(h = 10\), \(T = 39.7\), \(\epsilon = 0.01/T\), and \(d_u = d_v = d_w = 0.1\) for the Goodwin model, and \(T = 7\) for the FitzHugh–Nagumo model. The star symbols indicate parameter combinations for which $\min_{\gamma}\mathrm{CV}[y_1] > \min_{\gamma}\mathrm{CV}[y_2] > \min_{\gamma}\mathrm{CV}[y_3]$.
For the Goodwin model, these simulations used \(\alpha_1 = \alpha_2 = \alpha_3 = 1\), \(\beta = 1\), \(h = 10\), \(T = 39.7\), \(\epsilon = 0.01/T\), and \(D = 1.0 \times 10^{-5}\), while \(k\) and \(K\) were varied. For the FitzHugh–Nagumo model, parameters were fixed at \(a = 0.5\), \(b = 1.0\), \(c = -0.5\), \(\alpha = 1\), \(\beta = 1\), \(T = 7\), \(\epsilon = 0.01/T\), and \(D = 1.0 \times 10^{-5}\).

\paragraph*{Fig. 3.}
\label{Fig. 3}
{\bf Minimum CV across networks with different connection topologies.}
Noise behavior across different network wiring patterns. Displayed CV values represent the mean across 100 repetitions (see Methods).
\textbf{A:} (left) Eight network wiring patterns considered in this study.
(right-top) Period variability for the eight network wiring patterns at a constant degradation rate for the Goodwin, FitzHugh–Nagumo, and phase models.
(right-bottom) Output CV as a function of the shortest distance from the clock (number of edges).
Simulations were performed with the Goodwin clock parameters: $D = 0.1 \times 10^{-5}$, $\epsilon = 0.01/T$, $h = 10$, $k = 1$, $K = 100$, $T = 39.7$, $d_u = d_v = d_w = 0.1$, and output parameters $\alpha = 1$, $\beta = 1$, and $\gamma = 12.59$.
For the FitzHugh–Nagumo model, the parameter values were set to $D = 1.0 \times 10^{-4}$, $a = 0.5$, $b = 1.0$, and $c = -0.5$, with output parameters $\alpha = 1$, $\beta = 1$, and $\gamma = 10.0$.
For the phase model, simulations were performed with $\omega = 2\pi$, $D = 1.0 \times 10^{-2}$, $\epsilon = 0.01$, $\alpha = 1$, $\beta = 1$, and $\gamma = 10.0$.
\textbf{B:} (left) Structure of a complex network (1 clock, 75 outputs).
(right) Output CVs in the network. Node color distinguishes the clock from output nodes, and node shape indicates distance from the clock.
Simulations were performed with the Goodwin model. The parameter values were set to $D = 1.0 \times 10^{-5}$, $h = 10$, $k = 1$, $K = 1$, $T = 39.7$, $\epsilon = 0.01/T$, $d_u = d_v = d_w = 0.1$, and output parameters $\alpha = 1$, $\beta = 1$, $\gamma = 10.0$. 
\textbf{C:} (left) Structure of a cascade network (1 clock, 15 outputs).
(right) CV as a function of distance from the clock.
Simulations used the same Goodwin clock parameters as in B, with output parameters $\alpha = 1$, 
$\beta = 11.5$, and $\gamma = 10$.
\clearpage

\bibliography{references.bib}

@book{delvecchio2014biomolecular,
  author    = {Domitilla Del Vecchio and Richard M. Murray},
  title     = {Biomolecular Feedback Systems},
  publisher = {Princeton University Press},
  address   = {Princeton, NJ},
  year      = {2014},
  isbn       = {9780691161532}
}

@article{Son2008,
  author  = {Son, Gi Hoon and Chung, Soyoung and Choe, Hyeon Kyeong and Kim, Hyo-Dong and Baik, Seung-Mi and Lee, Hyeok and Lee, Kyu Young and Sun, Woo-Jin and Kim, Hyunho and Cho, Sunghoon and others},
  title   = {Adrenal peripheral clock controls the autonomous circadian rhythm of glucocorticoid by causing rhythmic steroid production},
  journal = {Proceedings of the National Academy of Sciences},
  year    = {2008},
  volume  = {105},
  number  = {52},
  pages   = {20970--20975},
  doi     = {10.1073/pnas.0806962105}
}

@article{Cho2012,
  author  = {Cho, Hyun Kook and Zhao, Xiping and Hatori, Masako and Yu, Richard T. and Barish, Grant D. and Lam, Mark T.Y. and Chong, Lee-Way and DiTacchio, Luca and Atkins, Andrew R. and Glass, Christopher K. and Liddle, Christopher and Auwerx, Johan and Downes, Michael and Panda, Satchidananda and Evans, Ronald M.},
  title   = {Regulation of circadian behaviour and metabolism by REV-ERB-$\alpha$ and REV-ERB-$\beta$},
  journal = {Nature},
  year    = {2012},
  volume  = {485},
  pages   = {123--127},
  doi     = {10.1038/nature11048}
}

@article{Markson2013,
  author  = {Markson, Jonathan S. and Piechura, Jessica R. and Puszynska, Anna M. and O'Shea, Erin K.},
  title   = {Circadian Control of Global Gene Expression by the Cyanobacterial Master Regulator RpaA},
  journal = {Cell},
  year    = {2013},
  volume  = {155},
  number  = {6},
  pages   = {1396--1408},
  doi     = {10.1016/j.cell.2013.11.005}
}

@incollection{Johnson:2004book,
author={Johnson, CH and Elliott, J and Foster, R and Honma, KI and Kronauer, R},
title={Fundamental properties of circadian rhythms},
booktitle={Chronobiology: biological timekeeping},
editor={Dunlap, JC and Loros, JJ and DeCoursey, PJ},
publisher={Sinauer Associates},
address={Sunderland, MA},
pages={67--105},
year={2004},
chapter={3}
}

@article{micklem2021cut,
  title={Cut the noise or couple up: Coordinating circadian and synthetic clocks},
  author={Micklem, Chris N and Locke, James CW},
  journal={iScience},
  volume={24},
  number={9},
  year={2021},
  publisher={Elsevier},
  doi={10.1016/j.isci.2021.103051}
}

@article{kaji2023enhanced,
  title={Enhanced precision of circadian rhythm by output system},
  author={Kaji, Hotaka and Mori, Fumito and Ito, Hiroshi},
  journal={Journal of Theoretical Biology},
  volume={574},
  pages={111621},
  year={2023},
  publisher={Elsevier},
  doi={https://doi.org/10.1016/j.jtbi.2023.111621}
}

@article{gonze2002robustness,
  title={Robustness of circadian rhythms with respect to molecular noise},
  author={Gonze, Didier and Halloy, Jos{\'e} and Goldbeter, Albert},
  journal={Proceedings of the National Academy of Sciences},
  volume={99},
  number={2},
  pages={673--678},
  year={2002},
  publisher={National Acad Sciences},
  doi={10.1073/pnas.022628299}
}

@article{kori2012structure,
  title={Structure of cell networks critically determines oscillation regularity},
  author={Kori, Hiroshi and Kawamura, Yoji and Masuda, Naoki},
  journal={Journal of Theoretical Biology},
  volume={297},
  pages={61--72},
  year={2012},
  publisher={Elsevier},
  doi={https://doi.org/10.1016/j.jtbi.2011.12.007}
}

@article{mori2013period,
  title={Period variability of coupled noisy oscillators},
  author={Mori, Fumito and Kori, Hiroshi},
  journal={Physical Review E},
  volume={87},
  number={3},
  pages={030901},
  year={2013},
  publisher={APS},
  doi={https://doi.org/10.1103/PhysRevE.87.030901}
}

@article{mori2022noninvasive,
  title={Noninvasive inference methods for interaction and noise intensities of coupled oscillators using only spike time data},
  author={Mori, Fumito and Kori, Hiroshi},
  journal={Proceedings of the National Academy of Sciences},
  volume={119},
  number={6},
  pages={e2113620119},
  year={2022},
  publisher={National Acad Sciences},
  doi={https://doi.org/10.1073/pnas.211362011}
}

@article{mori2016precision,
  title={Precision of collective oscillations in complex dynamical systems with noise},
  author={Mori, Fumito and Mikhailov, Alexander S},
  journal={Physical Review E},
  volume={93},
  number={6},
  pages={062206},
  year={2016},
  publisher={APS},
  doi={https://doi.org/10.1103/PhysRevE.93.062206}
}

@article{nagoshi2004circadian,
  title={Circadian gene expression in individual fibroblasts: cell-autonomous and self-sustained oscillators pass time to daughter cells},
  author={Nagoshi, Emi and Saini, Camille and Bauer, Christoph and Laroche, Thierry and Naef, Felix and Schibler, Ueli},
  journal={Cell},
  volume={119},
  number={5},
  pages={693--705},
  year={2004},
  publisher={Elsevier},
  doi={10.1016/j.cell.2004.11.015}
}

@article{mihalcescu2004resilient,
  title={Resilient circadian oscillator revealed in individual cyanobacteria},
  author={Mihalcescu, Irina and Hsing, Weihong and Leibler, Stanislas},
  journal={Nature},
  volume={430},
  number={6995},
  pages={81--85},
  year={2004},
  publisher={Nature Publishing Group UK London},
  doi={https://doi.org/10.1038/nature02533}
}

@article{nakamura2022adaptive,
  title={Adaptive diversification in the cellular circadian behavior of Arabidopsis leaf-and root-derived cells},
  author={Nakamura, Shunji and Oyama, Tokitaka},
  journal={Plant and Cell Physiology},
  volume={63},
  number={3},
  pages={421--432},
  year={2022},
  publisher={Oxford University Press UK},
  doi={10.1101/2021.06.10.447857}
}

@article{li2020noise,
  title={Noise-driven cellular heterogeneity in circadian periodicity},
  author={Li, Yan and Shan, Yongli and Desai, Ravi V and Cox, Kimberly H and Weinberger, Leor S and Takahashi, Joseph S},
  journal={Proceedings of the National Academy of Sciences},
  volume={117},
  number={19},
  pages={10350--10356},
  year={2020},
  publisher={National Academy of Sciences},
  doi={https://doi.org/10.1073/pnas.1922388117}
}

@article{gould2018coordination,
  title={Coordination of robust single cell rhythms in the Arabidopsis circadian clock via spatial waves of gene expression},
  author={Gould, Peter D and Domijan, Mirela and Greenwood, Mark and Tokuda, Isao T and Rees, Hannah and Kozma-Bognar, Laszlo and Hall, Anthony JW and Locke, James CW},
  journal={Elife},
  volume={7},
  pages={e31700},
  year={2018},
  publisher={eLife Sciences Publications, Ltd},
  doi={https://doi.org/10.7554/eLife.31700}
}

@article{nishino2013transcription,
  title={Transcription fluctuation effects on biochemical oscillations},
  author={Nishino, Ryota and Sakaue, Takahiro and Nakanishi, Hiizu},
  journal={PLoS One},
  volume={8},
  number={4},
  pages={e60938},
  year={2013},
  publisher={Public Library of Science San Francisco, USA},
  doi={https://doi.org/10.1371/journal.pone.0060938}
}

@article{chabot2007stochastic,
  title={Stochastic gene expression out-of-steady-state in the cyanobacterial circadian clock},
  author={Chabot, Jeffrey R and Pedraza, Juan M and Luitel, Prashant and Van Oudenaarden, Alexander},
  journal={Nature},
  volume={450},
  number={7173},
  pages={1249--1252},
  year={2007},
  publisher={Nature Publishing Group UK London},
  doi={https://doi.org/10.1038/nature06395}
}

@article{goodwin1965oscillatory,
  title={Oscillatory behavior in enzymatic control processes},
  author={Goodwin, Brian C},
  journal={Advances in Enzyme Regulation},
  volume={3},
  pages={425--437},
  year={1965},
  publisher={Elsevier},
  doi={https://doi.org/10.1016/0065-2571(65)90067-1}
}

@article{mangan2003structure,
  title={Structure and function of the feed-forward loop network motif},
  author={Mangan, Shmoolik and Alon, Uri},
  journal={Proceedings of the National Academy of Sciences},
  volume={100},
  number={21},
  pages={11980--11985},
  year={2003},
  publisher={National Academy of Sciences},
  doi={https://doi.org/10.1073/pnas.2133841100}
}

@article{dublanche2006noise,
  title={Noise in transcription negative feedback loops: simulation and experimental analysis},
  author={Dublanche, Yann and Michalodimitrakis, Konstantinos and K{\"u}mmerer, Nico and Foglierini, Mathilde and Serrano, Luis},
  journal={Molecular Systems Biology},
  volume={2},
  number={1},
  pages={MSB4100081},
  year={2006},
  publisher={Springer},
  doi={https://doi.org/10.1038/msb4100081}
}

@article{chalancon2012interplay,
  title={Interplay between gene expression noise and regulatory network architecture},
  author={Chalancon, Guilhem and Ravarani, Charles NJ and Balaji, S and Martinez-Arias, Alfonso and Aravind, L and Jothi, Raja and Babu, M Madan},
  journal={Trends in Genetics},
  volume={28},
  number={5},
  pages={221--232},
  year={2012},
  publisher={Elsevier},
  doi={10.1016/j.tig.2012.01.006 }
}

@article{becskei2000engineering,
  title={Engineering stability in gene networks by autoregulation},
  author={Becskei, Attila and Serrano, Luis},
  journal={Nature},
  volume={405},
  number={6786},
  pages={590--593},
  year={2000},
  publisher={Nature Publishing Group UK London},
  doi={https://doi.org/10.1038/35014651}
}

@article{dibner2010mammalian,
  title={The mammalian circadian timing system: organization and coordination of central and peripheral clocks},
  author={Dibner, Charna and Schibler, Ueli and Albrecht, Urs},
  journal={Annual Review of Physiology},
  volume={72},
  number={1},
  pages={517--549},
  year={2010},
  publisher={Annual Reviews},
  doi={https://doi.org/10.1146/annurev-physiol-021909-135821}
}

@article{mohawk2012central,
  title={Central and peripheral circadian clocks in mammals},
  author={Mohawk, Jennifer A and Green, Carla B and Takahashi, Joseph S},
  journal={Annual Review of Neuroscience},
  volume={35},
  pages={445--462},
  year={2012},
  publisher={Annual Reviews},
  doi={https://doi.org/10.1146/annurev-neuro-060909-153128}
}

@article{takahashi2017transcriptional,
  title={Transcriptional architecture of the mammalian circadian clock},
  author={Takahashi, Joseph S},
  journal={Nature Reviews Genetics},
  volume={18},
  number={3},
  pages={164--179},
  year={2017},
  publisher={Nature Publishing Group UK London},
  doi={https://doi.org/10.1038/nrg.2016.150}
}

@article{kurosawa2002comparative,
  title={Comparative study of circadian clock models, in search of processes promoting oscillation},
  author={Kurosawa, Gen and Mochizuki, Atsushi and Iwasa, Yoh},
  journal={Journal of Theoretical Biology},
  volume={216},
  number={2},
  pages={193--208},
  year={2002},
  publisher={Elsevier},
  doi={https://doi.org/10.1006/jtbi.2002.2546}
}

@book{guckenheimer2013nonlinear,
  title={Nonlinear oscillations, dynamical systems, and bifurcations of vector fields},
  author={Guckenheimer, John and Holmes, Philip},
  year={2013},
  publisher={Springer Science \& Business Media}
}

@article{kaji2025sinusoidal,
  title={Sinusoidal regulation reduces circadian period variability},
  author={Kaji, Hotaka and Mori, Fumito and Maruyama, Osamu and Ito, Hiroshi},
  journal={Scientific reports},
  volume={15},
  number={1},
  pages={33843},
  year={2025},
  publisher={Nature Publishing Group UK London},
  doi={https://doi.org/10.1038/s41598-025-04614-z}
}

@article{xiao2025low,
  title={Low temperature abolishes human cellular circadian rhythm through Hopf bifurcation},
  author={Xiao, Yaoyao and Sainoo, Yuko and Nishimura, Takayuki and Ito, Hiroshi},
  journal={npj Systems Biology and Applications},
  volume={12},
  number={1},
  pages={5},
  year={2025},
  publisher={Nature Publishing Group},
  doi={https://doi.org/10.1038/s41540-025-00628-5}
}

@article{murayama2017low,
  title={Low temperature nullifies the circadian clock in cyanobacteria through Hopf bifurcation},
  author={Murayama, Yoriko and Kori, Hiroshi and Oshima, Chiaki and Kondo, Takao and Iwasaki, Hideo and Ito, Hiroshi},
  journal={Proceedings of the National Academy of Sciences},
  volume={114},
  number={22},
  pages={5641--5646},
  year={2017},
  publisher={National Academy of Sciences},
  doi={https://doi.org/10.1073/pnas.1620378114}
}

\includepdf[pages=-]{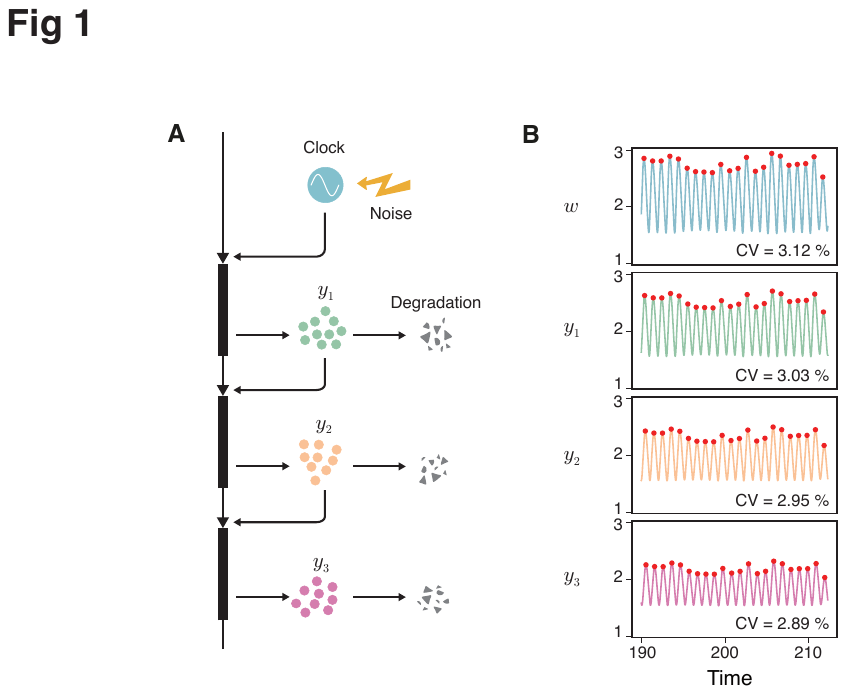}
\includepdf[pages=-]{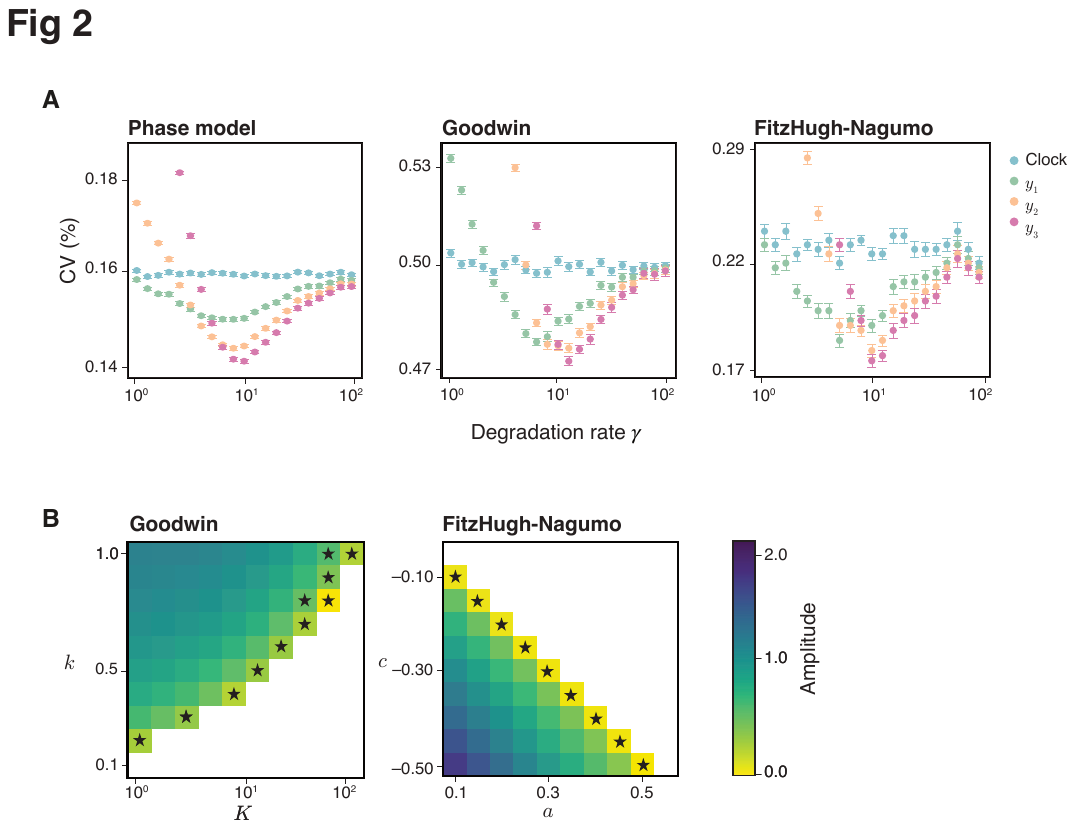}
\includepdf[pages=-]{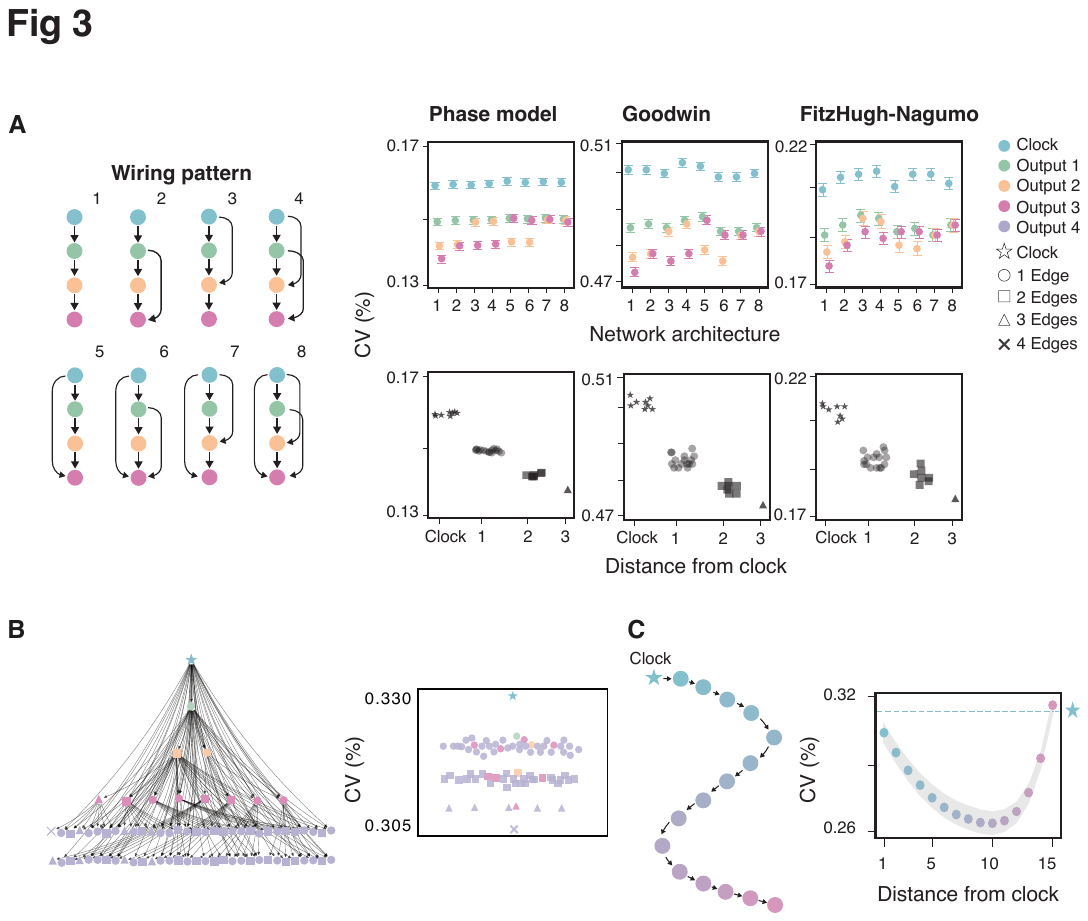}
\includepdf[pages=-]{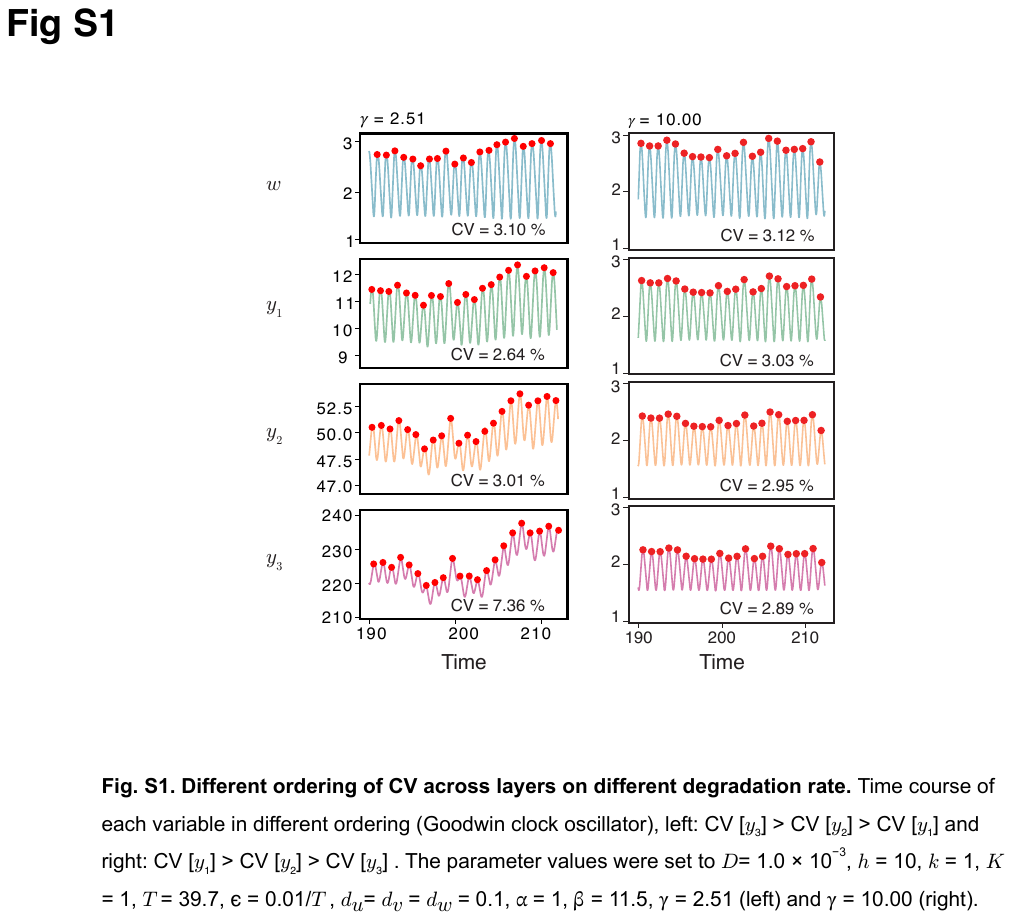}
\includepdf[pages=-]{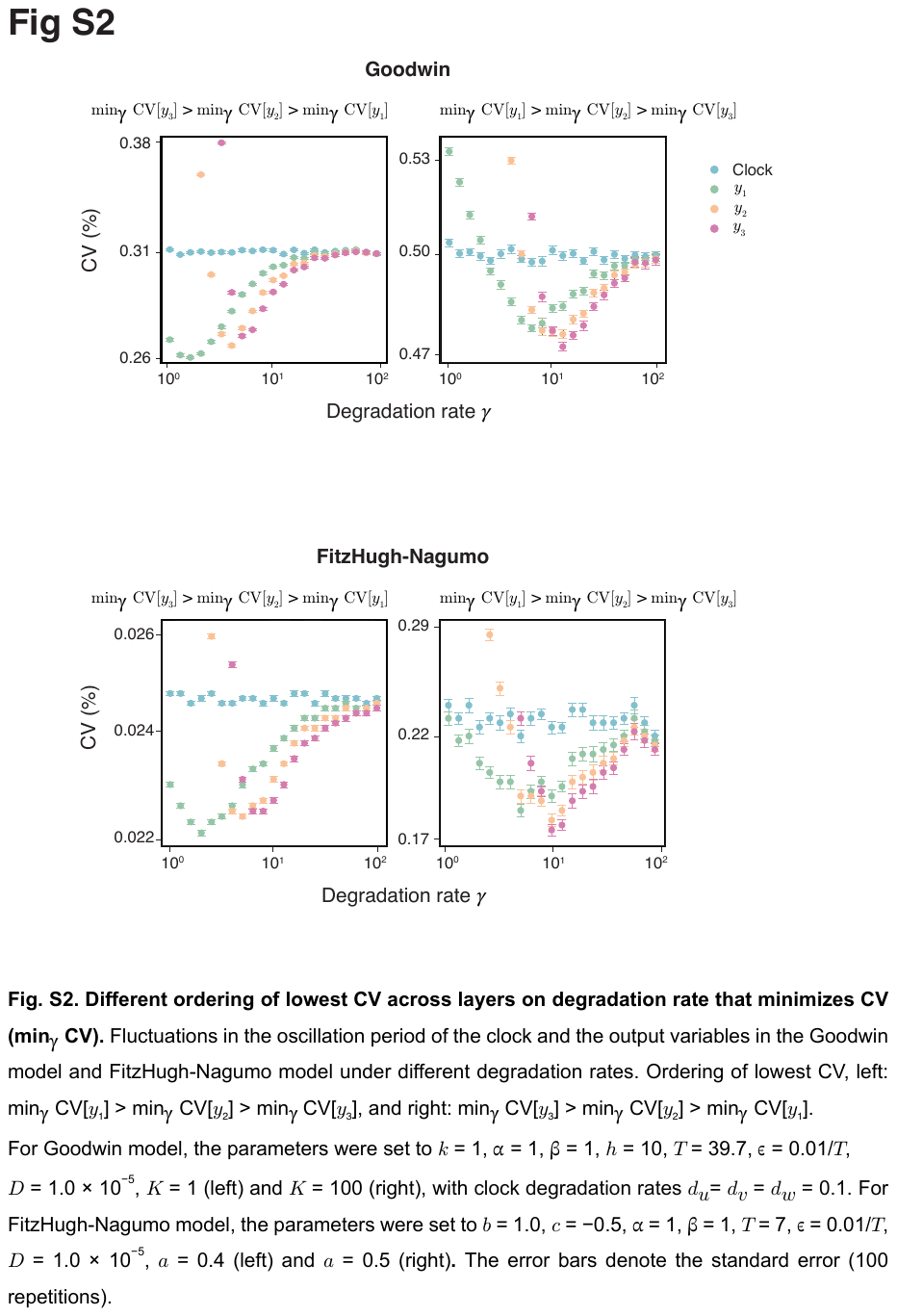}
\includepdf[pages=-]{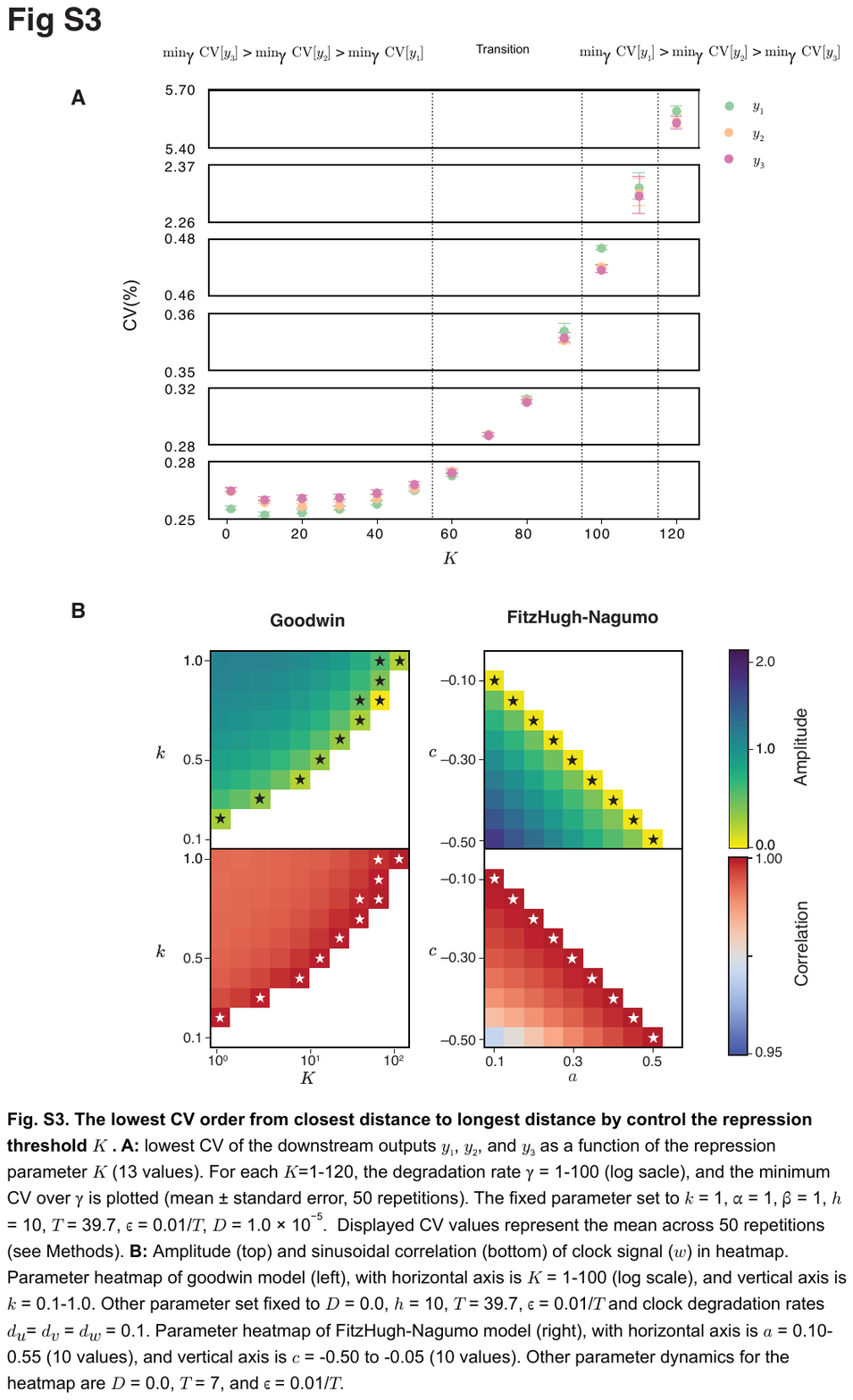}
\includepdf[pages=-]{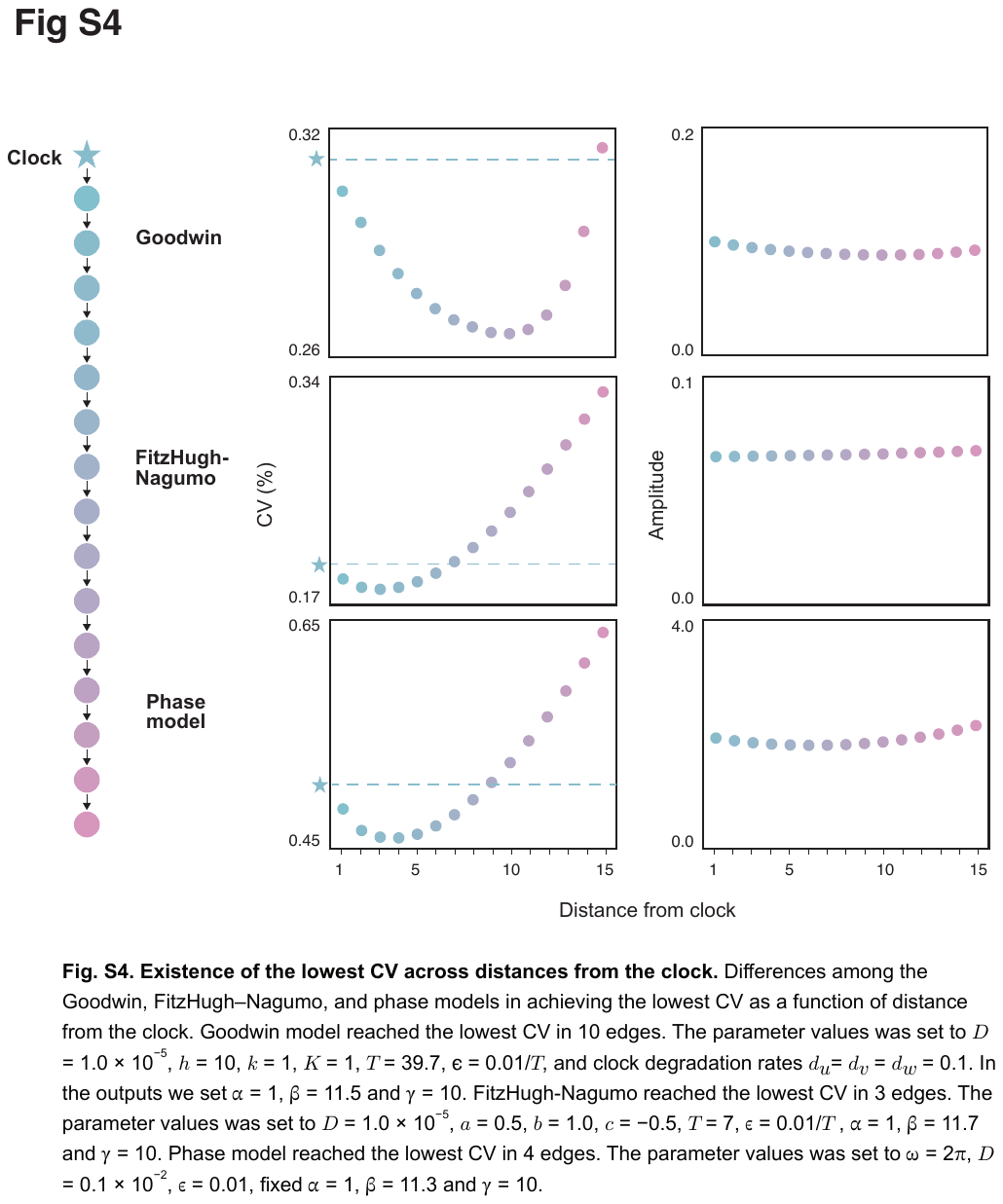}

\end{document}